\documentclass[a4paper,twocolumn,showpacs,floatfix,prd,nofootinbib]{revtex4-1}
\usepackage{graphicx}
\usepackage{epsf}
\usepackage{epsfig}
\usepackage{amssymb,amsmath}
\usepackage[usenames]{color}
\usepackage{times}
\usepackage[colorlinks=true,linktocpage=false,citecolor=blue,linkcolor=blue,urlcolor=blue]{hyperref}
\usepackage{enumerate}
\usepackage{booktabs,dcolumn}
\usepackage{pstricks}
\usepackage{rotating}
\newcolumntype{d}[1]{D{.}{.}{#1}}
\def\th{\theta}
\def\nn{\nonumber}
\def\be{\begin{equation}}
\def\ee{\end{equation}}
\def\beq{\begin{eqnarray}}
\def\eeq{\end{eqnarray}}
\def\bi{\begin{itemize}}
\def\ei{\end{itemize}}
\def\ben{\begin{enumerate}}
\def\een{\end{enumerate}}

\begin{document}

\author{Francesco Pannarale}%
\email{francesco.pannarale@ligo.org}%
\affiliation{%
  School of Physics and Astronomy, Cardiff University, The Parade,
  Cardiff CF24 3AA, United Kingdom\\
  Max-Planck-Institut f{\"u}r Gravitationsphysik, Albert Einstein
  Institut, Potsdam 14476, Germany%
}

\title{\bf Black hole remnant of black hole-neutron star coalescing
  binaries with arbitrary black hole spin}

\begin{abstract}
  A model for determining the dimensionless spin parameter and mass of
  the black hole remnant of black hole-neutron star mergers with
  arbitrary initial black hole spin angular momentum, binary mass
  ratio, and neutron star mass and cold equation of state is
  formulated. Tests against numerical-relativity results are carried
  out, showing that both the dimensionless spin parameter and the
  final mass are accurately reproduced. For the first time, the
  behavior of both quantities and of the $l=2$, $m=2$, $n=0$
  quasinormal mode frequency is inspected throughout the parameter
  space. Predictions of this frequency may be exploited to guide
  gravitational-wave modeling and detection efforts and to extract
  physical information from detected gravitational-wave signals that
  would help us break degeneracies between binary black hole and black
  hole-neutron star systems, improve our understanding of compact
  binary formation, and constrain the neutron star equation of state.
\end{abstract}

\maketitle

\section{Introduction}
Following its formation, a black hole-neutron star (BH-NS) binary
system loses orbital energy and angular momentum by emitting
gravitational radiation. During the inspiral stage, it readjusts its
orbital separation accordingly, until the energy reservoir is emptied
and the two compact objects merge. The remnant of this coalescence
consists of a black hole possibly surrounded by a hot, massive
accretion torus~\cite{ShibataTaniguchilrr-2011-6}. BH-NS binaries have
not been observed yet, but population synthesis studies suggest that
BH-NS mergers are likely to occur frequently in the Hubble
volume~\cite{Kalogera2007, Belczynski07, Belczynski08,
  Oshaughnessy2008}. By detecting the radiation emitted during their
inspiral and merger, gravitational-wave (GW) observatories --- such as
LIGO~\cite{ligowebpage}, Virgo~\cite{virgowebpage},
KAGRA~\cite{kagrawebpage}, and the Einstein
Telescope~\cite{Punturo:2010} --- are therefore likely to be involved
in the first direct observations of BH-NS systems. Advanced LIGO is
expected to see between $0.2$ and $300$ BH-NS binaries per
year~\cite{Abadie:2010}. Other than being GW
sources~\cite{Hannam2013a,Lundgren2013b,Harry2013,Oshaughnessy2013},
BH-NS binaries, or mixed binaries, are among promising progenitor
candidates for a fraction of the short-hard gamma-ray bursts (SGRBs)
we observe~\cite{Nakar:2007yr, Berger2011}. Further, NSs in BH-NS
systems may undergo tidal distortions and disruptions. This aspect has
drawn considerable attention in relation to the possibility of
constraining the equation of state (EOS) of matter at supranuclear
densities~\cite{Vallisneri00, Ferrari:2009bw, Pannarale2011,
  Lackey2013} and in relation to the physical processes the merger
ejecta may be involved in. Radioactive decays triggered in the
outflows by the formation of heavy isotopes may power day-long optical
transients (``kilonov\ae''), and shocks generated by the outflows
themselves in an interstellar medium of sufficiently high density may
also lead to visible electromagnetic counterparts~\cite{Roberts2011,
  Metzger2012}. Additionally, the ejecta are believed to be linked to
the abundances of heavy elements formed in $r$-processes, by rapid
neutron capture~\cite{Lattimer74}.

Numerical-relativity dynamical simulations of BH-NS mergers will
ultimately allow us to fully understand and explore the physics behind
these events. Research with this intent was initiated fairly
recently~\cite{Janka99a,Ruffert99b, Kluzniak99c, Rosswog05,
  Loeffler06a, Sopuerta:2006bw, Shibata06d, ShibataUryu:2007,
  Etienne2007b, ShibataTaniguchi2008, Rantsiou2008, Duez:2008rb,
  Etienne:2008re, Shibata:2009cn, Duez09, Kyutoku2010, Chawla:2010sw,
  Foucart2010, Kyutoku2011, Kyutoku2011err, Foucart2011, Lackey2012,
  Etienne2012, Shibata2012err, Etienne2012b, Deaton2013, Foucart2013a,
  Kyutoku2013, Foucart2013b, Lovelace2013, Paschalidis2013b} and has
benefitted from relativistic quasiequilibrium
studies~\cite{Taniguchi05, Grandclement06, Faber06a, Faber05,
  Tsokaros2007, Taniguchi07, Taniguchi:2008a}. As fully relativistic
dynamical simulations remain challenging and computationally
intensive, pseudo-Newtonian calculations, e.g.~\cite{Ruffert2010}, and
analytical studies, e.g.~\cite{Shibata96, Wiggins00, Vallisneri00,
  Berti08, Hanna2008, Ferrari09, Ferrari:2009bw, Damour:2009wj,
  Pannarale2010, Pannarale2011, Foucart2012, Pannarale2012}, have been
carried out as well for BH-NS systems. These studies generally focus
on a few selected physical aspects --- possibly even a single one ---
of BH-NS mergers and provide insight on the problem by offering access
to the large space of parameters of mixed binaries at low
computational costs.

This paper belongs to this last category of approaches and extends the
work presented in~\cite{Pannarale2012} (paper I, henceforth). Its
focus is to predict the dimensionless spin parameter and the mass of
the BH remnant of BH-NS coalescing binaries. As opposed to paper I, no
restriction to cases with parallel BH spin angular momentum and
orbital angular momentum is made: in this sense, this article
generalizes the model of paper I. Our semianalytical approach is based
on the model of Buonanno, Kidder, and Lehner (BKL) to estimate the
final BH spin of a BH-BH coalescence which exploits angular momentum
and energy conservation for the binary from its last stable orbit
onwards~\cite{Buonanno:07b}. In order to account for the fraction of
orbital angular momentum and energy that may not accrete immediately
onto the BH, our model requires us to know the baryonic mass
$M_\text{b,torus}$ surviving the BH-NS merger and surrounding the BH
remnant. As in paper I, when making predictions for BH-NS systems for
which no numerical-relativity simulations exist yet, we rely on the
relativistic fitting formula of~\cite{Foucart2012} to determine
$M_\text{b,torus}$. This fit was calibrated to numerical-relativity
simulations of aligned BH-NS mergers, but, as recently pointed
out~\cite{Stone2013,Foucart2013a}, it may also be used for mixed
binaries with nonparallel BH spin angular momentum and orbital angular
momentum. In terms of $M_\text{b,torus}$, a BH-NS merger with
misaligned BH spin is roughly equivalent to a BH-NS merger with an
effective, lower, aligned BH spin angular momentum given by
Eq.\,(\ref{eq:aEff}). This allows us to go beyond the formulation of
paper I, and to generalize the model formulated therein. The key
equations of our approach are
Eqs.\,(\ref{eq:model-tilted-1})-(\ref{eq:model-tilted-Mf}), along with
the ansatz of Eq.\,(\ref{eq:fbridge}), which appeared in paper I and
is left unchanged for simplicity. Our tests show that our simple
method accurately reproduces the results of numerical-relativity
simulations of BH-NS mergers with tilted initial BH
spin~\cite{Foucart2010,Foucart2013a}, despite the mathematical
complexity of the BH-NS coalescence problem. Further, by exploring the
BH-NS space of parameters, we illustrate how our predictions may be
used to guide GW detection efforts and to extract valuable
astrophysical information from detected GW signals.

The paper is organized as follows. In Sec.\,\ref{sec:orig-model} we
review the approach developed in paper I to predict the spin parameter
and mass of BH remnants of BH-NS binaries with aligned BH spin angular
momentum and orbital angular momentum. In Sec.\,\ref{sec:new-model} we
propose an extension of this method and successfully test it against
available numerical-relativity data. In Sec.\,\ref{sec:results} we
gather and discuss the results obtained by systematically varying the
binary mass ratio, the initial BH spin parameter and inclination, and
the NS mass and EOS. Finally, in Sec.\,\ref{sec:conclusions}, we draw
our conclusions and collect our remarks.

\section{Nonprecessing BH-NS Binaries}\label{sec:orig-model}
In paper I we formulated and validated an approach to estimate the
final spin parameter and mass of BH remnants of BH-NS mergers in which
the initial BH spin and the orbital angular momentum are aligned. Our
method was based on the BKL approach to estimate the final spin of
BH-BH mergers~\cite{Buonanno:07b} and is reviewed in this section,
prior to extending it to arbitrary initial BH spin orientations in the
next section.

We begin by recalling that we set the initial spin angular momentum of
the NS to zero, as this is believed to be a reliable approximation of
astrophysically realistic systems~\cite{Bildsten92,Kochanek92} and
because the same assumption is made in building initial data for
current BH-NS merger numerical simulations, which form our test
bed. Let us now collect some expressions for equatorial orbits around
a Kerr BH of dimensionless spin parameter $a$ and mass
$m$~\cite{Bardeen72}. The orbital separation at the innermost stable
circular orbit (ISCO) is given by
\begin{align}
  \label{eq:rISCO}
  \bar{r}_\text{ISCO} &= [3+Z_2\mp\sqrt{(3-Z_1)(3+Z_1+2Z_2)}]\nn\\
  Z_1 &= 1 +
  (1-a^2)^{1/3}\left[(1+a)^{1/3}+(1-a)^{1/3}\right]\nn\\
  Z_2 &= \sqrt{3a^2+Z_1^2}\,,
\end{align}
where the upper(lower) sign holds for corotating(counterrotating)
orbits, and where $\bar{r}=r/m$ denotes the (dimensionless)
Boyer-Lindquist radial coordinate. In this same notation, the orbital
angular momentum per unit mass of a test particle orbiting the BH may
be expressed as
\begin{align}
\label{eq:lz}
l_z(\bar{r},a) &= \pm \frac{\bar{r}^2\mp
  2a\sqrt{\bar{r}}+a^2}{\sqrt{\bar{r}}(\bar{r}^2-3\bar{r}\pm
  2a\sqrt{\bar{r}})^{1/2}}\,,
\end{align}
while its orbital energy reads
\begin{align}
\label{eq:e}
e(\bar{r},a) &= \frac{\bar{r}^2-2\bar{r}\pm
  a\sqrt{\bar{r}}}{\bar{r}(\bar{r}^2-3\bar{r}\pm
    2a\sqrt{\bar{r}})^{1/2}}\,.
\end{align}

Given a BH-NS binary formed by a BH of mass (at isolation)
$M_\text{BH}$ and initial dimensionless spin parameter $a_\text{i}$,
and a NS of mass (at isolation) $M_\text{NS}$, the dimensionless spin
parameter $a_\text{f}$ of the BH remnant may be determined by solving
(numerically) the following equation:
\begin{widetext}
\begin{align}
\label{eq:model-eq-af}
a_\text{f} = \frac{a_\text{i}M_\text{BH}^2 +
  l_z(\bar{r}_\text{ISCO,f},a_\text{f})M_\text{BH}\{[1-f(\nu)]M_\text{NS}+f(\nu)M_\text{b,NS}-M_\text{b,torus}\}}{[M\left\{1-[1-e(\bar{r}_\text{ISCO,i},a_\text{i})]\nu\right\}-e(\bar{r}_\text{ISCO,f},a_\text{f})M_\text{b,torus}]^2}\,.
\end{align}
\end{widetext}
In the previous expression, the notation $M=M_\text{BH}+M_\text{NS}$
was used, $\bar{r}_\text{ISCO,i}$ and $\bar{r}_\text{ISCO,f}$ denote
$\bar{r}_\text{ISCO}$ calculated for the initial and the final BH spin
parameter, respectively, $\nu=M_\text{BH}M_\text{NS}/M^2$ is the
symmetric mass ratio, $M_\text{b,torus}$ is the baryonic mass of the
torus remnant, and
\begin{align}
\label{eq:fbridge}
f(\nu) = \left\{
\begin{array}{ll}
  0 & \nu \leq 0.16 \\
  \frac{1}{2}\Big[1-\cos\Big(\frac{\pi(\nu - 0.16)}{2/9-0.16}\Big)\Big] & 0.16<\nu<2/9 \\
  1 & 2/9\leq\nu\leq 0.25 \\
\end{array}
\right.\nn\\
\end{align}
is a function that regulates the transition between nondisruptive and
disruptive coalescences, which happen for low and high values of
$\nu$, respectively. We remind the reader that
Eq.\,(\ref{eq:model-eq-af}) follows from angular momentum and energy
conservation considerations. The terms
$-l_z(\bar{r}_\text{ISCO,f},a_\text{f})M_\text{BH}M_\text{b,torus}$
and $-e(\bar{r}_\text{ISCO,f},a_\text{f})M_\text{b,torus}$ that appear
in the numerator and denominator, respectively, are expressions that
approximate the angular momentum and energy of the NS material that
survives the merger, remains outside the BH horizon, and is not
accreted onto the BH. Let us also remark that the term
$-M[1-e(\bar{r}_\text{ISCO,i},a_\text{i})]\nu$ in the denominator
accounts for the orbital energy radiated away in GWs by the binary
system during its inspiral~\cite{Barausse2012b}.

The closed expression in Eq.\,(\ref{eq:model-eq-af}) may be solved
numerically with root-finding techniques to determine the
dimensionless spin parameter of the BH remnant. Once this is done, the
denominator on the right-hand side of Eq.\,(\ref{eq:model-eq-af})
automatically provides a prediction for the mass of the BH remnant
itself:
\begin{align}
\label{eq:model-eq-Mf}
M_\text{f} &=
M\left\{1-[1-e(\bar{r}_\text{ISCO,i},a_\text{i})]\nu\right\}-e(\bar{r}_\text{ISCO,f},a_\text{f})M_\text{b,torus}\,.
\end{align}
%

\section{Generic Initial BH Spin Configurations}\label{sec:new-model}
Consider now the more general scenario in which the initial spin
angular momentum of the BH has an arbitrary direction and is therefore
not necessarily parallel to the orbital angular momentum of the
binary: in order to predict the spin and mass of the BH remnant of the
BH-NS coalescence, it is possible to track the BKL approach and to
modify our model ---
Eqs.\,(\ref{eq:model-eq-af})-(\ref{eq:model-eq-Mf}) --- accordingly.

We stress that we do not drop the assumption that the NS initial spin
angular momentum vanishes. Once again, this approximation is present
in the numerical-relativity simulations against which our model is
tested. The total spin of the system thus coincides with the BH spin
angular momentum $\vec{S}_\text{BH}$ and has magnitude
$S_\text{BH}\equiv |\vec{S}_\text{BH}|$ and direction
$\hat{S}_\text{BH}\equiv \vec{S}_\text{BH}/S_\text{BH}$. For an
unequal mass binary with a single spinning component, i.e.~the case we
are treating, the magnitude of the total spin and the angle
$\th_\text{i}\equiv \arccos(\hat{L}_\text{orb}\cdot
\hat{S}_\text{BH})$ between the total spin and the orbital angular
momentum, having direction $\hat{L}_\text{orb}$, are constant up to
the innermost stable spherical orbit (ISSO), to an excellent
approximation~\cite{Apostolatos94}.

As in the BKL approach, we assume that $M_\text{BH}$, $M_\text{NS}$,
$\vec{S}_\text{BH}$, and $\hat{L}_\text{orb}$ are all known at some
point of the inspiral prior to the ISSO. Under the assumption that the
angular momentum of the system changes only by a small amount during
the merger and ringdown stages (cf.~paper I), the total angular
momentum $\vec{J}_\text{f}\equiv\vec{a}_\text{f}M_\text{f}^2$ of the
BH remnant may be computed by adding the initial spin angular momentum
of the system, $\vec{S}_\text{BH}$, and the orbital angular momentum
accreted onto the BH itself. As pointed out in~\cite{Buonanno:07b},
the key point in estimating the orbital angular momentum contribution
is to allow for the last stable orbit to be inclined with respect to
the total angular momentum of the BH remnant
$\vec{J}_\text{f}$. Bearing this in mind, we generalize the term in
Eq.\,(\ref{eq:model-eq-af}) for the orbital angular momentum accreted
onto the BH as
\be
\label{eq:LorbAccr}
\vec{l}_\text{orb}(\bar{r}_\text{ISSO,f},a_\text{f},\th_\text{f})M_\text{BH}\{[1-f(\nu)]M_\text{NS}+f(\nu)M_\text{b,NS}-M_\text{b,torus}\}\,,
\ee
where $\vec{l}_\text{orb}$ is the orbital angular momentum per unit
mass of a test particle orbiting a BH \emph{at the ISSO} and
$\th_\text{f}$ labels the angle between $\vec{l}_\text{orb}$ at the
ISSO and $\vec{J}_\text{f}$. The angular momentum of the BH remnant is
therefore the sum of the last expression and
$\vec{S}_\text{BH}\equiv\vec{a}_\text{i}M_\text{BH}^2$. We can project
this sum along $\vec{J}_\text{f}$ itself and orthogonally to it. This
readily yields the two equations
\begin{widetext}
\begin{align}
\label{eq:model-tilted-1}
l_\text{orb}(\bar{r}_\text{ISSO,f},a_\text{f},\th_\text{f})M_\text{BH}\{[1-f(\nu)]M_\text{NS}+f(\nu)M_\text{b,NS}-M_\text{b,torus}\}\cos\th_\text{f}
+ a_\text{i}M_\text{BH}^2\cos(\th_\text{i}-\th_\text{f}) &= a_\text{f}M_\text{f}^2\\
\label{eq:model-tilted-2}
l_\text{orb}(\bar{r}_\text{ISSO,f},a_\text{f},\th_\text{f})M_\text{BH}\{[1-f(\nu)]M_\text{NS}+f(\nu)M_\text{b,NS}-M_\text{b,torus}\}\sin\th_\text{f}
- a_\text{i}M_\text{BH}^2\sin(\th_\text{i}-\th_\text{f}) &= 0\,,
\end{align}
\end{widetext}
where $l_\text{orb}=|\vec{l}_\text{orb}|$,
$a_\text{i,f}=|\vec{a}_\text{i,f}|$, and the mass of the BH remnant
$M_\text{f}$ is
\begin{align}
\label{eq:model-tilted-Mf}
M_\text{f} &=\nn\\
&M\left\{1-[1-e(\bar{r}_\text{ISSO,i},a_\text{i},\th_\text{i})]\nu\right\}-e(\bar{r}_\text{ISSO,f},a_\text{f},\th_\text{f})M_\text{b,torus}\,,
\end{align}
a straightforward generalization of Eq.\,(\ref{eq:model-eq-Mf}). In
extending the equatorial case model of paper I,
Eqs.\,(\ref{eq:model-eq-af})-(\ref{eq:model-eq-Mf}), we decide not to
modify the ansatz of Eq.\,(\ref{eq:fbridge}) that defines $f(\nu)$:
this is done for the sake of simplicity and because this does not seem
to be necessary in light of the validation tests performed (see
Sec.\,\ref{sec:tests}). The formulation of the model is completed by
discussing $l_\text{orb}$, $e$, and $M_\text{b,torus}$. As
in~\cite{Buonanno:07b}, we compute the orbital angular momentum
$l_\text{orb}$ of the inclined orbit using the fitting formula
of~\cite{Hughes:2002ei}; similarly, we choose to compute the orbital
energy $e$ at the ISSO via the equivalent fitting formula provided in
the same reference. The two expressions are
\begin{align}
  \label{eq:lISSO}
  l_\text{orb}(\bar{r}_\text{ISSO},a,\th) &=
  \frac{1}{2}(1+\cos\th)l_z^\text{pro}(\bar{r}_\text{ISCO}^\text{pro},a)\nn\\
  &+\frac{1}{2}(1-\cos\th)|l_z^\text{ret}(\bar{r}_\text{ISCO}^\text{ret},a)|\\
  \label{eq:eISSO}
  e(\bar{r}_\text{ISSO},a,\th) &=
  \frac{1}{2}(1+\cos\th)e^\text{pro}(\bar{r}_\text{ISCO}^\text{pro},a)\nn\\
  &+\frac{1}{2}(1-\cos\th)|e^\text{ret}(\bar{r}_\text{ISCO}^\text{ret},a)|\,,
\end{align}
where $\bar{r}_\text{ISCO}^\text{pro,ret}$, $l_z^\text{pro,ret}$, and
$e^\text{pro,ret}$ are given in Eqs.\,(\ref{eq:rISCO})-(\ref{eq:e})
for prograde ($\text{pro}$) and retrograde ($\text{ret}$)
orbits. Finally, the baryonic mass surviving the merger,
$M_\text{b,torus}$, may be estimated by following the consideration
made by Stone and collaborators in~\cite{Stone2013} and further
validated by Foucart and collaborators in~\cite{Foucart2013a}: in
terms of $M_\text{b,torus}$, a BH-NS merger with misaligned BH spin is
roughly equivalent to a BH-NS merger with lower, aligned BH spin
angular momentum with dimensionless parameter $a_\text{i,Eff}$. We
determine this effective dimensionless spin parameter by solving
\begin{align}
  \label{eq:aEff}
  \bar{r}_\text{ISCO}(a_\text{i,Eff}) &= \bar{r}_\text{ISSO}(a_\text{i},\th_\text{i}) \nn\\
  &=\frac{1}{2}(1+\cos\th_\text{i})\bar{r}_\text{ISCO}^\text{pro}(a_\text{i})\nn\\
  &+\frac{1}{2}(1-\cos\th_\text{i})\bar{r}_\text{ISCO}^\text{ret}(a_\text{i})\,,
\end{align}
where $\bar{r}_\text{ISCO}(a_\text{i,Eff})$ may be calculated via
Eq.\,(\ref{eq:rISCO}) and where the second equality expresses
$\bar{r}_\text{ISSO}(a_\text{i},\th_\text{i})$ and is carried out by
following the fitting formulas of~\cite{Hughes:2002ei}, in accordance
with Eqs.\,(\ref{eq:lISSO})-(\ref{eq:eISSO}). $M_\text{b,torus}$ is
then computed using $a_\text{i,Eff}$ in the relativistic fitting
formula of~\cite{Foucart2012} for aligned BH-NS mergers.

Summarizing, given a BH-NS binary formed by a NS with gravitational
mass $M_\text{NS}$, baryonic mass $M_\text{b,NS}$, and radius
$R_\text{NS}$ (determined by its EOS), and a BH with gravitational
mass $M_\text{BH}$, and initial spin of (dimensionless) magnitude
$a_\text{i}$ and initial inclination $\th_\text{i}$ with respect to
the orbital angular momentum vector of the binary, the properties of
the BH remnant may be calculated as follows:
\ben[~~~~(1)]
\item solve Eq.\,(\ref{eq:aEff}) numerically for $a_\text{i,Eff}$;
\item use $a_\text{i,Eff}$ to calculate the baryonic mass of the
  matter that survives the merger, $M_\text{b,torus}$,
  following~\cite{Foucart2012};
\item solve Eqs.\,(\ref{eq:model-tilted-1})-(\ref{eq:model-tilted-2})
  numerically to determine (a) the (dimensionless) magnitude of the
  spin angular momentum of the BH remnant, $a_\text{f}$, and (b) its
  inclination angle $\th_\text{f}$ with respect to the orbital angular
  momentum;
\item determine the gravitational mass of the BH remnant by plugging
  $M_\text{b,torus}$ and $a_\text{f}$ in
  Eq.\,(\ref{eq:model-tilted-Mf}).
\een
As a final remark, we stress that for $\theta_\text{i}=0$, i.e.~for
nontilted BH spins, this procedure reduces to the model formulated and
tested in paper I.

\subsection{Validation of the model}\label{sec:tests}
Numerical-relativity simulations of BH-NS coalescences with misaligned
BH spin and orbital angular momentum vectors were reported
in~\cite{Foucart2010,Foucart2013a}, for a total of seven
binaries. These form the test bed currently available for our
model.\footnote{Sixteen BH-BH tests were carried out in the BKL
  paper~\cite{Buonanno:07b}.} The tests are collected in Table
\ref{tab:tests}. The first column of the table numbers the tests,
while the second column provides the reference in which the
numerical-relativity simulation was presented. The following three
columns are the physical parameters of the binary:
$Q=M_\text{BH}/M_\text{NS}$ (the binary mass ratio), $a_\text{i}$, and
$\theta_\text{i}$. In all cases, the NS is spinless, it is governed by
a $\Gamma=2$ polytropic EOS, and it has compactness
$C=M_\text{NS}/R_\text{NS}=0.144$. Columns six and eight provide the
numerical-relativity prediction for the magnitude of the dimensionless
spin angular momentum of the BH remnant, $a_\text{f}^\text{NR}$, and
its mass in units of $M$, $M_\text{f}^\text{NR}/M$. In columns seven
and nine we compare the estimates of our model, $a_\text{f}$ and
$M_\text{f}/M$, to $a_\text{f}^\text{NR}$ and
$M_\text{f}^\text{NR}/M$: in the former case we report the difference
$a_\text{f}-a_\text{f}^\text{NR}$, in the latter case the relative
error $\epsilon(M_\text{f}/M)$. In the last column we collect our
results for $\th_\text{f}$. Obviously the low number of test cases
does not allow us to test the model thoroughly, throughout the space
of parameters. It is noteworthy, however, that $Q$ takes both a low
and a high value, that $a_\text{i}$ reaches the relatively high value
$0.9$, and that $\theta_\text{i}$ runs up to $80^\circ$.

In paper I we established that we were able to reproduce the
numerical-relativity results for BH-NS binaries with equatorial orbits
with an absolute error $\Delta a_\text{f}\simeq 0.02$ on the
dimensionless spin parameter, and a relative error of $\sim 1$\% on
$M_\text{f}/M$, with a few test cases reaching a $\sim 2$\%
error. Table \ref{tab:tests} shows these standards are respected when
extending our phenomenological model to the case of binaries with a
misaligned BH spin. Even though the number of tests is low, and large
portions of the parameter space remain unexplored, the results are
strikingly positive, and they provide first evidence that the
generalization of the model of paper I formulated in this paper works
successfully.

The angle $\theta_\text{f}$ (last columns of Table \ref{tab:tests}) is
not linked to results of numerical-relativity simulations in an
immediate way. Given its definition, measuring this angle in a fully
relativistic simulation is not a well posed task. However,
$\theta_\text{f}$ is a useful quantity as it provides an estimate for
the angle $\psi_\text{d}$ between the disk possibly formed in a BH-NS
merger and the BH remnant spin angular momentum
vector~\cite{Stone2013}. The interest in $\psi_\text{d}$ derives from
its direct link to the possibility of having quasiperiodic signals of
jet precession as an observational signature of SGRBs originating from
BH-NS mergers, as opposed to ones originating from binary NS
systems~\cite{Stone2013}. A unique $\psi_\text{d}$ angle is an
idealized concept, as the postmerger dynamics of BH-NS systems are
complex, but a disk tilt angle measurement is provided for the
simulations reported in~\cite{Foucart2010}. This is calculated where
the surface density of the disk is maximum, at late times. The best we
can do to link $\theta_\text{f}$ to numerical-relativity results is,
therefore, to compare it to the quantity just discussed. For the first
four cases discussed in Table \ref{tab:tests}, Foucart and
collaborators find tilt angles of $2^\circ$, $4^\circ$, $7^\circ$, and
$8^\circ$~\cite{Foucart2010}. These are all a factor $\sim 3$-$4$
smaller than $\theta_\text{f}$. In light of these comparisons, we may
thus conclude that $\theta_\text{f}$ may be used to constrain the disk
tilt angle via
\begin{equation}
  \label{eq:disk-tilt-constraints}
  \theta_\text{f}/4 \lesssim \psi_\text{d} \lesssim \theta_\text{f}/3\,.
\end{equation}

\begin{table}
  \caption{\label{tab:tests} Tests against results of numerical-relativity simulations of BH-NS coalescences with misaligned spin reported in the literature. Each row is a test case numbered by the index in the first column. The remaining columns provide the reference in which the simulation is discussed, the binary mass ratio ($Q$), the initial magnitude of the dimensionless spin angular momentum of the BH ($a_\text{i}$), the angle it forms with respect to the orbital angular momentum ($\theta_\text{i}$), the numerical-relativity prediction for the magnitude of the dimensionless spin of the BH remnant ($a_\text{f}^\text{NR}$), the difference between our estimate $a_\text{f}$ and $a_\text{f}^\text{NR}$, the numerical-relativity prediction for the final gravitational mass in units of $M=M_\text{BH}+M_\text{NS}$ ($M_\text{f}^\text{NR}/M$), the relative difference between our estimate for $M_\text{f}/M$ and $M_\text{f}^\text{NR}/M$, and our prediction for the direction of the spin angular momentum of the BH remnant ($\theta_\text{f}$). In each simulation the NS is irrotational, has compactness $C=0.144$, and is governed by a $\Gamma=2$ polytropic EOS.}
  \begin{tabular}{l@{\hspace{0.25cm}}c@{\hspace{0.25cm}}c@{\hspace{0.25cm}}c@{\hspace{0.25cm}}c@{\hspace{0.25cm}}c@{\hspace{0.15cm}}d{2.2}@{\hspace{0.15cm}}c@{\hspace{0.15cm}}d{2.2}@{\hspace{0.15cm}}c}
    \toprule[1.pt]
    \toprule[1.pt]
    \addlinespace[0.3em]
    & Ref. & $Q$ & $a_\text{i}$ & $\theta_\text{i}$ & $a_\text{f}^\text{NR}$  & \multicolumn{1}{c}{$a_\text{f}-a_\text{f}^\text{NR}$} & $M_\text{f}^\text{NR}/M$ & \multicolumn{1}{c}{$\epsilon(M_\text{f}/M)$} & $\theta_\text{f}$ \\
    \addlinespace[0.2em]
    \midrule[1.pt]
    \addlinespace[0.2em]
    1  & \cite{Foucart2010} & $3$ & $0.5$ & $20^\circ$ & $0.76$ & -0.01 & $0.95$ &  0.00 & $8^\circ$ \\
    2  & \cite{Foucart2010} & $3$ & $0.5$ & $40^\circ$ & $0.74$ & -0.01 & $0.96$ &  0.01 & $16^\circ$\\
    3  & \cite{Foucart2010} & $3$ & $0.5$ & $60^\circ$ & $0.71$ & -0.02 & $0.96$ &  0.00 & $22^\circ$\\
    4  & \cite{Foucart2010} & $3$ & $0.5$ & $80^\circ$ & $0.66$ & -0.02 & $0.97$ &  0.00 & $27^\circ$\\
    \midrule
    5 & \cite{Foucart2013a} & $7$ & $0.9$ & $20^\circ$ & $0.909$ & 0.02 & $0.939$ & 0.02 & $16^\circ$\\
    6 & \cite{Foucart2013a} & $7$ & $0.9$ & $40^\circ$ & $0.898$ & 0.01 & $0.959$ & 0.01 & $31^\circ$\\
    7 & \cite{Foucart2013a} & $7$ & $0.9$ & $60^\circ$ & $0.862$ & 0.01 & $0.978$ & 0.01 & $45^\circ$\\
    \bottomrule[1.pt]
    \bottomrule[1.pt]
  \end{tabular}
\end{table}

In Table \ref{tab:tests-MtorusNR}, we repeat the tests of Table
\ref{tab:tests} using the numerical-relativity values of
$M_\text{b,torus}$, instead of the ones obtained by exploiting
Eq.\,(\ref{eq:aEff}) and the relativistic fit
of~\cite{Foucart2012}. As we see, this slightly improves the
performance of our model. More specifically, the relative error on
$M_\text{f}/M$ (eighth column) no longer exceeds $1$\%, and
$|a_\text{f}-a_\text{f}^\text{NR}|$ (sixth column) is $0.02$ in only
one case (as opposed to three). This is the same trend observed in the
tests of paper I, and seems to indicate that the fitting formula for
$M_\text{b,torus}$ is our main source of error. Further, and
importantly, we notice that the results for $\th_\text{f}$ (ninth and
last column) are unaffected, therefore confirming
Eq.\,(\ref{eq:disk-tilt-constraints}). Finally, the seventh column of
Table \ref{tab:tests-MtorusNR} shows the difference
$a_\text{f}^\text{BKL}-a_\text{f}^\text{NR}$ between the BKL BH-BH
prediction for the final spin parameter and the BH-NS
numerical-relativity result for the same quantity.\footnote{We remind
  the reader that $M_\text{f}=M$ in the BKL approach.} With one
exception, our model improves the original BKL estimate for
$a_\text{f}$. This conclusion also holds when comparing values for
$a_\text{f}^\text{BKL}-a_\text{f}^\text{NR}$ to values for
$a_\text{f}-a_\text{f}^\text{NR}$ from Table \ref{tab:tests}.

\begin{table}
  \caption{\label{tab:tests-MtorusNR} Tests against results of numerical-relativity simulations of BH-NS coalescences with misaligned spin reported in the literature. The first five columns are organized as in Table \ref{tab:tests}. The sixth column provides the difference between our estimate $a_\text{f}$ and $a_\text{f}^\text{NR}$: as opposed to Table \ref{tab:tests}, where the fit of~\cite{Foucart2012} and the approximation outlined in Eq.\,(\ref{eq:aEff}) were used,  $a_\text{f}$ is computed  using the numerical-relativity result for $M_\text{b,torus}$. In the seventh column, we report the difference between the BKL prediction for the final spin parameter and $a_\text{f}^\text{NR}$. The last two columns are the same as in Table \ref{tab:tests} (the values are once more obtained using the numerical-relativity result for $M_\text{b,torus}$).}
  \resizebox{\columnwidth}{!}{%
  \begin{tabular}{l@{\hspace{0.25cm}}c@{\hspace{0.25cm}}c@{\hspace{0.25cm}}c@{\hspace{0.25cm}}c@{\hspace{0.15cm}}d{2.2}@{\hspace{0.15cm}}d{2.2}@{\hspace{0.15cm}}c@{\hspace{0.15cm}}c}
    \toprule[1.pt]
    \toprule[1.pt]
    \addlinespace[0.3em]
    & Ref. & $Q$ & $a_\text{i}$ & $\theta_\text{i}$ & \multicolumn{1}{c}{$a_\text{f}-a_\text{f}^\text{NR}$} & \multicolumn{1}{c}{$a_\text{f}^\text{BKL}-a_\text{f}^\text{NR}$} & $\epsilon(M_\text{f}/M)$ & $\theta_\text{f}$ \\
    \addlinespace[0.2em]
    \midrule[1.pt]
    \addlinespace[0.2em]
    1  & \cite{Foucart2010} & $3$ & $0.5$ & $20^\circ$ & -0.01 & -0.02 & $0.00$ & $8^\circ$ \\
    2  & \cite{Foucart2010} & $3$ & $0.5$ & $40^\circ$ & -0.01 & -0.02 & $0.00$ & $16^\circ$\\
    3  & \cite{Foucart2010} & $3$ & $0.5$ & $60^\circ$ & -0.01 & -0.03 & $0.01$ & $22^\circ$\\
    4  & \cite{Foucart2010} & $3$ & $0.5$ & $80^\circ$ & -0.02 & -0.03 & $0.00$ & $27^\circ$\\
    \midrule
    5 & \cite{Foucart2013a} & $7$ & $0.9$ & $20^\circ$ & 0.01 & 0.00  & $0.01$ & $16^\circ$\\
    6 & \cite{Foucart2013a} & $7$ & $0.9$ & $40^\circ$ & 0.01 & -0.01 & $0.01$ & $31^\circ$\\
    7 & \cite{Foucart2013a} & $7$ & $0.9$ & $60^\circ$ & 0.01 & -0.01 & $0.00$ & $45^\circ$\\
    \bottomrule[1.pt]
    \bottomrule[1.pt]
  \end{tabular}}
\end{table}

In order to gauge how well our approach performs, and in addition to
the comparisons of Table \ref{tab:tests-MtorusNR}, in concluding this
section we would like to draw a comparison with a model formulated
in~\cite{Stone2013} to predict the spin of the BH remnant of BH-NS
mergers. We wish to remark that the scope of the work presented by
Stone and collaborators was \emph{not} to build the most accurate
model to calculate $\vec{a}_\text{f}$, but to study the origin of
SGRBs. The method of~\cite{Stone2013} is based on a
post-Newtonian fit, and it, too, uses the results
of~\cite{Foucart2012} to predict $M_\text{b,torus}$, albeit via the
Newtonian fit, and not the relativistic one used in this paper and in
paper I. The predictions for the magnitude of the dimensionless spin
angular momentum vector of the BH remnant yielded by the model
($a_\text{f}^\text{PN}$) are tested in Table II of~\cite{Stone2013}
against the numerical-relativity results
of~\cite{Foucart2010,Foucart2011,Foucart2013a}. In Table
\ref{tab:tests-Stone}, we report these same comparisons with the
addition of our predictions for $a_\text{f}$. The models seem to
perform equivalently for high BH masses, i.e.~for $Q=7$, regardless of
the initial spin configuration of the BH and the NS compactness. We
observe that, for these binaries, our prediction almost systematically
slightly exceeds the numerical-relativity result, while the
post-Newtonian model slightly underestimates it. For $Q=5$, that is,
for a moderate BH mass, the post-Newtonian model underestimates
$a_\text{f}^\text{NR}$ by $0.03$, while our model happens to reproduce
it exactly. The trend continues to progress this way as the mass ratio
is lowered: with the exception of cases $11$ and $17$, for which
$\{a_\text{i}=0.0, \th_\text{i}=0^\circ\}$ and $\{a_\text{i}=0.5,
\th_\text{i}=80^\circ\}$, our model performs significantly better. The
most notable case is the one with $\{a_\text{i}=0.9,
\th_\text{i}=0^\circ\}$, for which
$a_\text{f}^\text{NR}-a_\text{f}^\text{PN}=0.10$, while
$a_\text{f}^\text{NR}-a_\text{f}=-0.01$. From these comparisons, we
therefore conclude that, overall, our phenomenological model is better
at reproducing the numerical-relativity results.

\begin{table}
  \caption{\label{tab:tests-Stone} Comparison between the magnitude of the dimensionless spin angular momentum of the BH remnant of BH-NS coalescences, as predicted by numerical-relativity simulations ($a_\text{f}^\text{NR}$, seventh column), by the model of Stone and collaborators~\cite{Stone2013} ($a_\text{f}^\text{PN}$, eighth column), and our model ($a_\text{f}$, ninth column). The first five columns are organized as in Table \ref{tab:tests}. The sixth column specifies the compactness $C$ of the NS. In all cases the NS EOS is a $\Gamma=2$ polytrope, and the NS is irrotational. $a_\text{f}^\text{NR}$, $a_\text{f}^\text{PN}$, and $a_\text{f}$ are all reported with two significant figures.}
  \resizebox{\columnwidth}{!}{%
  \begin{tabular}{l@{\hspace{0.3cm}}c@{\hspace{0.3cm}}c@{\hspace{0.3cm}}c@{\hspace{0.35cm}}c@{\hspace{0.35cm}}c@{\hspace{0.4cm}}c@{\hspace{0.4cm}}c@{\hspace{0.4cm}}c}
    \toprule[1.pt]
    \toprule[1.pt]
    \addlinespace[0.3em]
    & Ref. & $Q$ & $a_\text{i}$ & $\theta_\text{i}$ & $C$ & $a_\text{f}^\text{NR}$ & $a_\text{f}^\text{PN}$ & $a_\text{f}$ \\
    \addlinespace[0.2em]
    \midrule[1.pt]
    \addlinespace[0.2em]
    1  & \cite{Foucart2011}  & $7$ & $0.5$ & $0^\circ$  & $0.144$ & $0.67$ & $0.66$ & $0.68$\\
    2  & \cite{Foucart2011}  & $7$ & $0.7$ & $0^\circ$  & $0.144$ & $0.80$ & $0.79$ & $0.81$\\
    3  & \cite{Foucart2011}  & $7$ & $0.9$ & $0^\circ$  & $0.144$ & $0.92$ & $0.91$ & $0.93$\\
    4  & \cite{Foucart2011}  & $5$ & $0.5$ & $0^\circ$  & $0.144$ & $0.71$ & $0.68$ & $0.71$\\
    \midrule
    5  & \cite{Foucart2013a} & $7$ & $0.9$ & $0^\circ$  & $0.170$ & $0.92$ & $0.91$ & $0.93$\\
    6  & \cite{Foucart2013a} & $7$ & $0.9$ & $0^\circ$  & $0.156$ & $0.92$ & $0.91$ & $0.93$\\
    7  & \cite{Foucart2013a} & $7$ & $0.9$ & $0^\circ$  & $0.144$ & $0.91$ & $0.91$ & $0.93$\\
    8  & \cite{Foucart2013a} & $7$ & $0.9$ & $20^\circ$ & $0.144$ & $0.91$ & $0.91$ & $0.93$\\
    9  & \cite{Foucart2013a} & $7$ & $0.9$ & $40^\circ$ & $0.144$ & $0.90$ & $0.90$ & $0.91$\\
    10 & \cite{Foucart2013a} & $7$ & $0.9$ & $60^\circ$ & $0.144$ & $0.86$ & $0.87$ & $0.87$\\
    \midrule
    11 & \cite{Foucart2010}  & $3$ & $0.0$ & $0^\circ$  & $0.144$ & $0.56$ & $0.54$ & $0.55$\\
    12 & \cite{Foucart2010}  & $3$ & $0.5$ & $0^\circ$  & $0.144$ & $0.77$ & $0.70$ & $0.76$\\
    13 & \cite{Foucart2010}  & $3$ & $0.9$ & $0^\circ$  & $0.144$ & $0.93$ & $0.83$ & $0.94$\\
    14 & \cite{Foucart2010}  & $3$ & $0.5$ & $20^\circ$ & $0.144$ & $0.76$ & $0.70$ & $0.75$\\
    15 & \cite{Foucart2010}  & $3$ & $0.5$ & $40^\circ$ & $0.144$ & $0.74$ & $0.69$ & $0.73$\\
    16 & \cite{Foucart2010}  & $3$ & $0.5$ & $60^\circ$ & $0.144$ & $0.71$ & $0.67$ & $0.69$\\
    17 & \cite{Foucart2010}  & $3$ & $0.5$ & $80^\circ$ & $0.144$ & $0.66$ & $0.64$ & $0.64$\\
    \bottomrule[1.pt]
    \bottomrule[1.pt]
  \end{tabular}}
\end{table}

\section{Results}\label{sec:results}
\begin{figure*}[htb]
  \begin{tabular*}{\textwidth}{c@{\extracolsep{\fill}}c}
  \includegraphics[scale=0.32,clip=true]{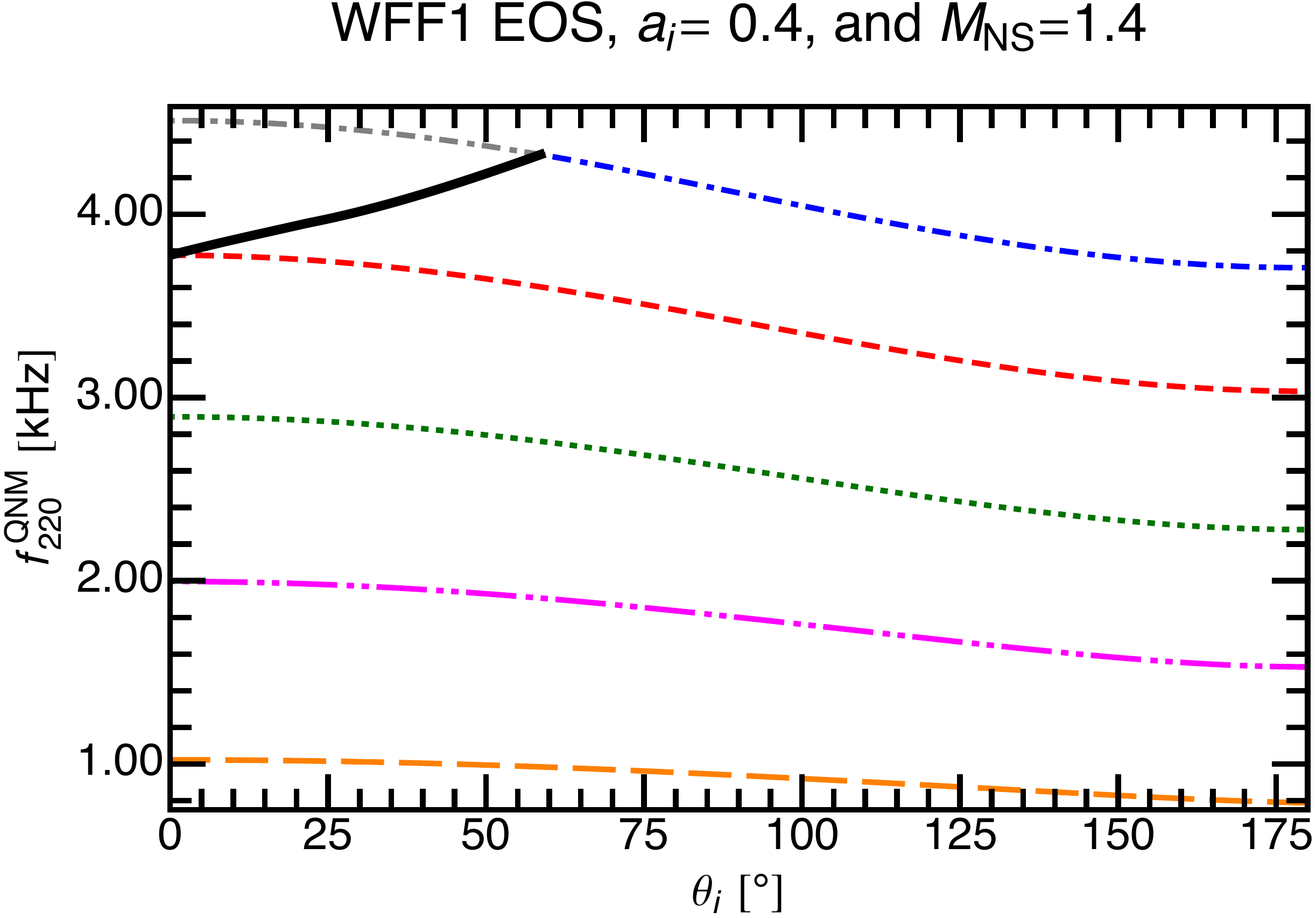}
  &
  \includegraphics[scale=0.32,clip=true]{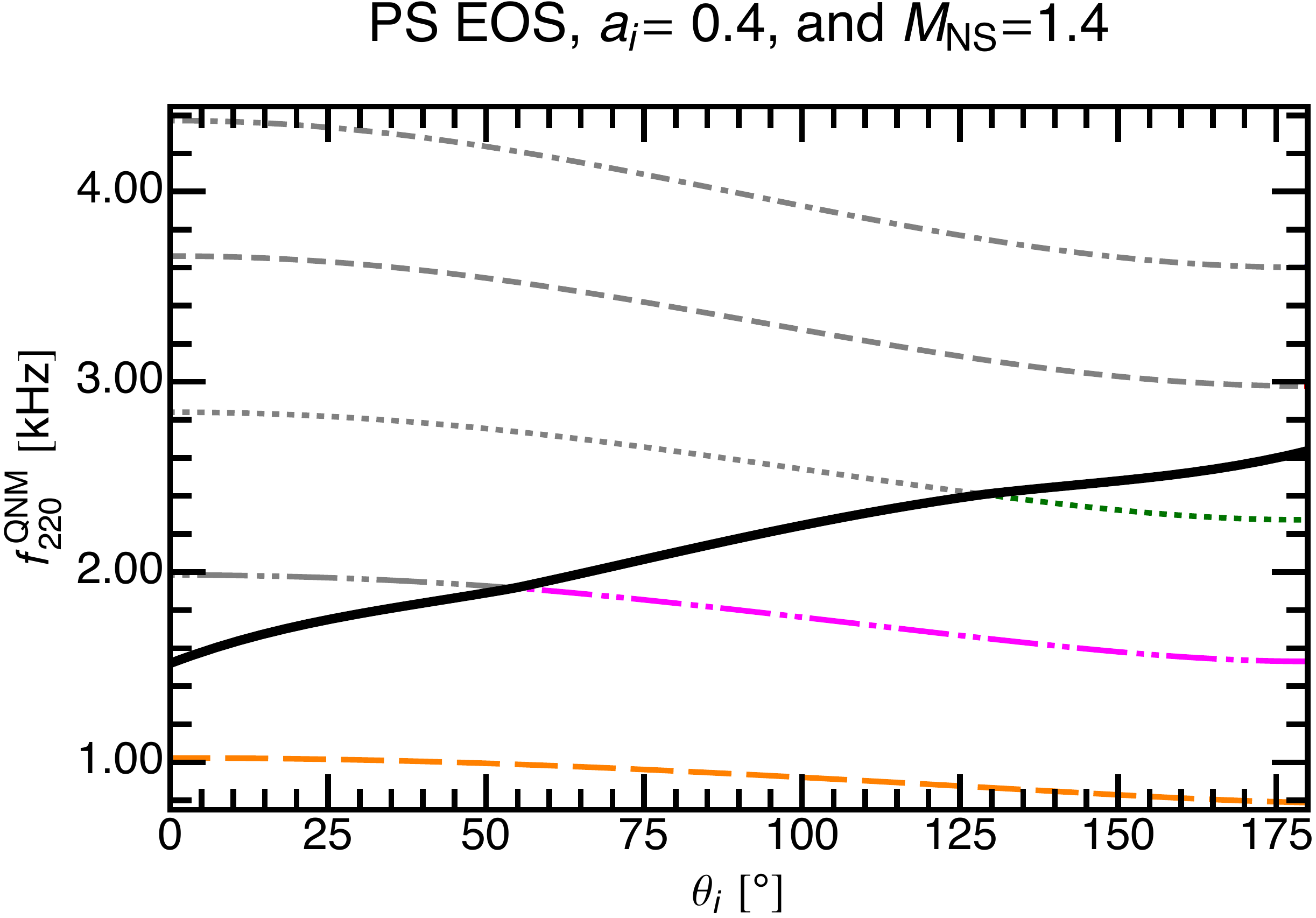}
  \\
  &
  \\
  \includegraphics[scale=0.32,clip=true]{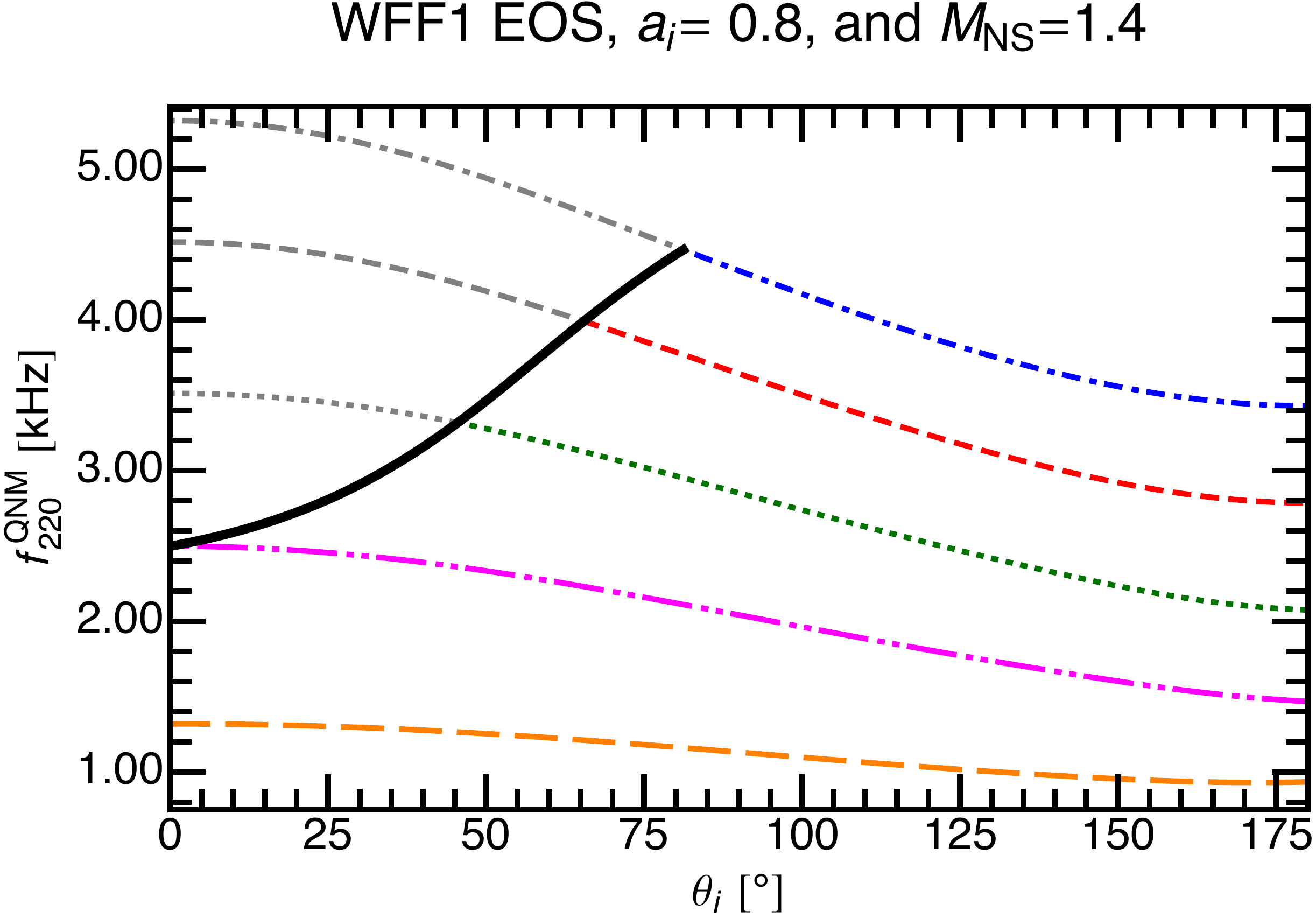}
  &
  \includegraphics[scale=0.32,clip=true]{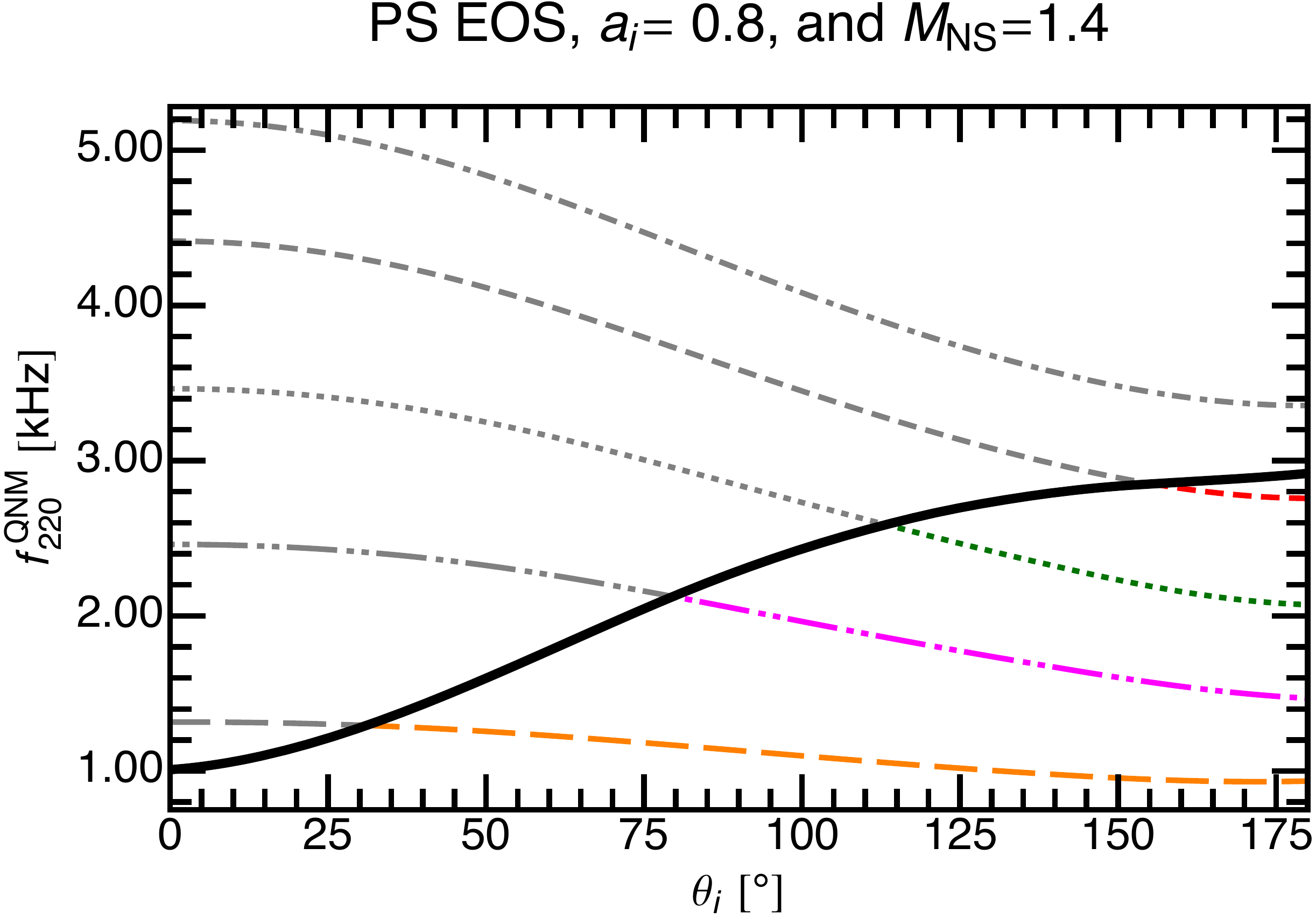}
  \\
  &
  \\
  \includegraphics[scale=0.32,clip=true]{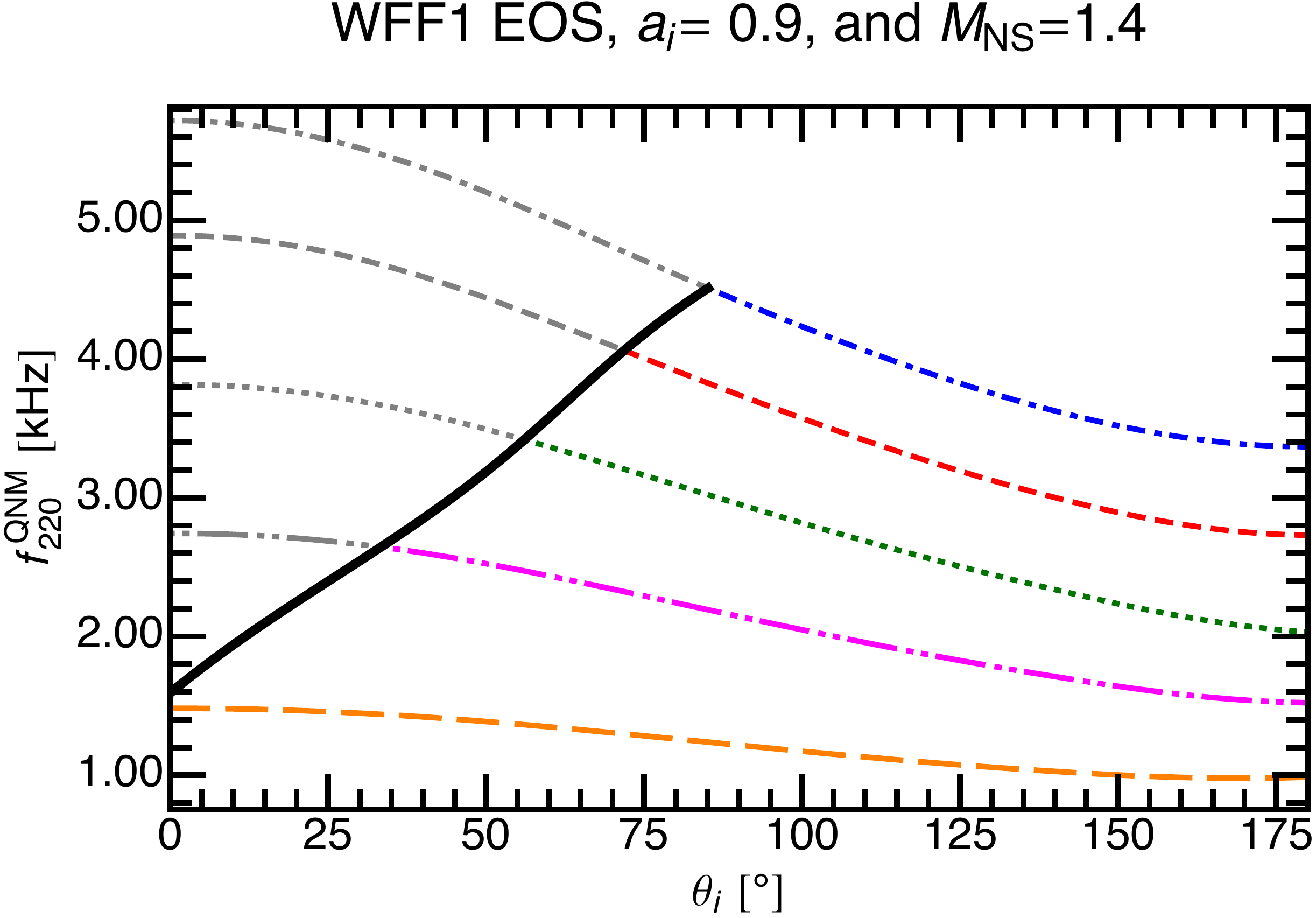}
  &
  \includegraphics[scale=0.32,clip=true]{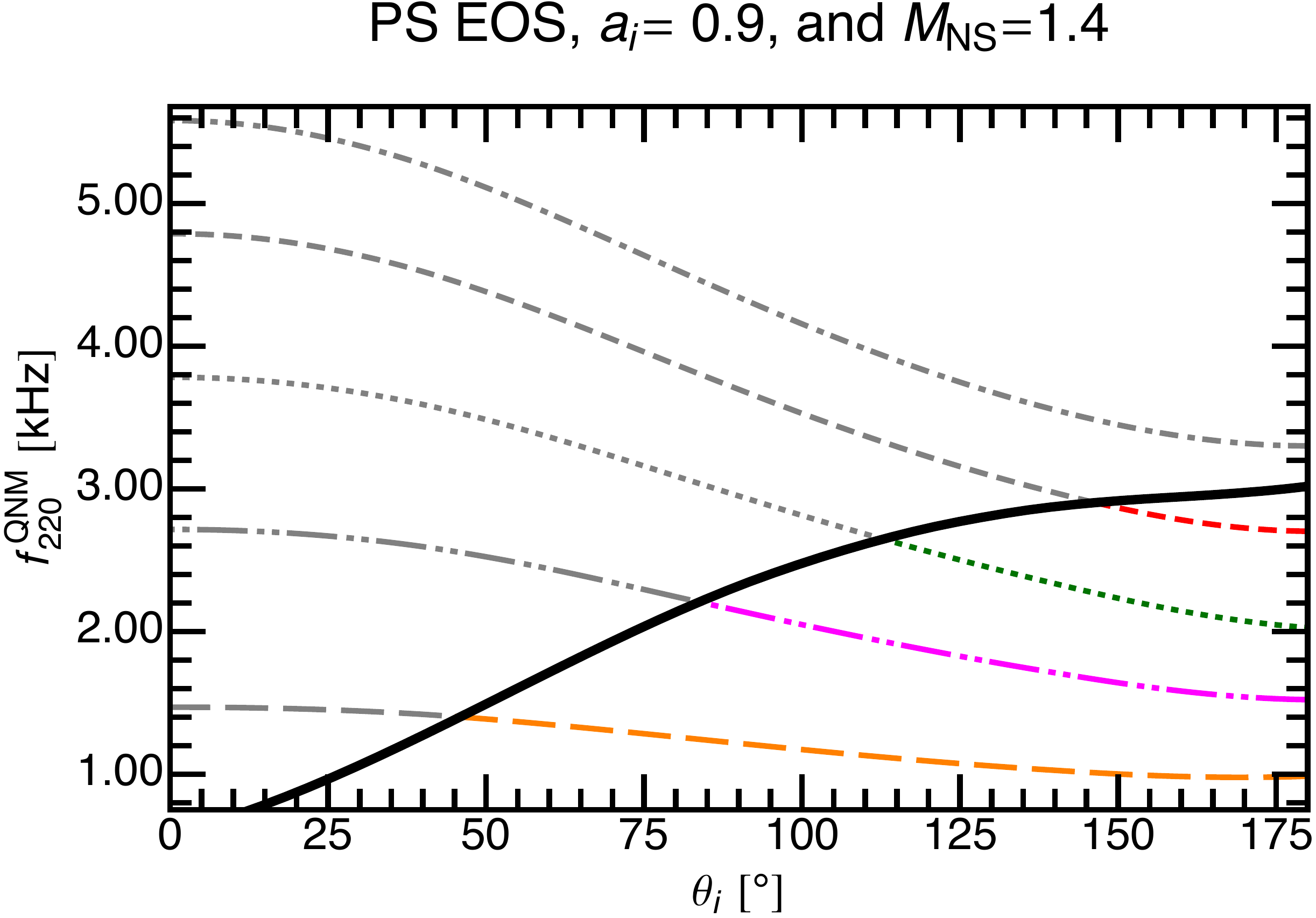}
  \\
  &
  \\
  \includegraphics[scale=0.32,clip=true]{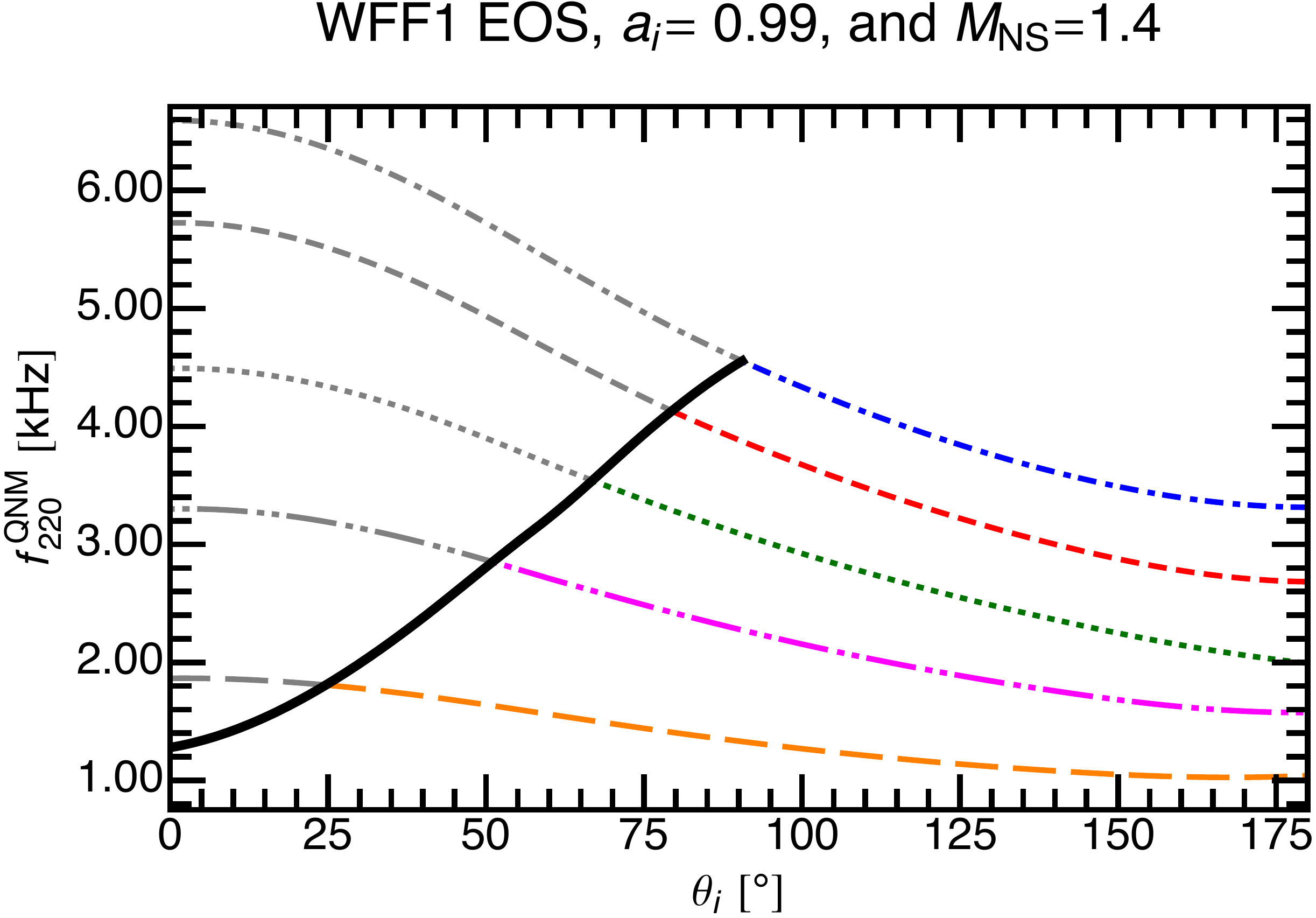}
  &  
  \includegraphics[scale=0.32,clip=true]{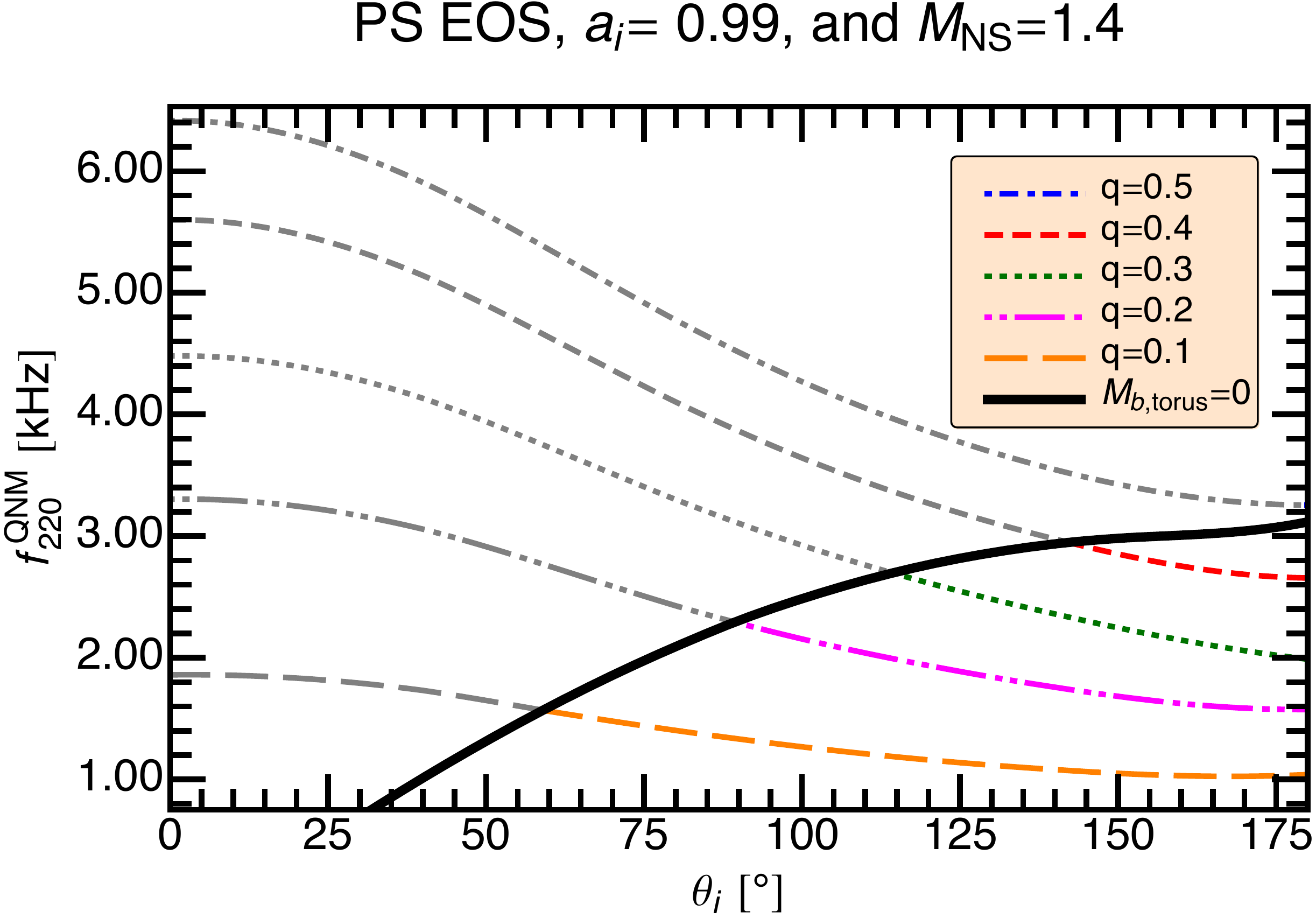}
  \end{tabular*}
  \caption{(Color online) $l=2$, $m=2$, $n=0$ quasinormal mode
    frequency, $f_{220}^\text{QNM}$, as a function of the initial tilt
    angle $\theta_\text{i}$. The NS mass is taken to be
    $1.4M_\odot$. Left and right panels refer to the WFF1 EOS and PS
    EOS, respectively. The initial BH spin parameter $a_\text{i}$
    increases from top to bottom, as indicated by the title of each
    panel. Different colors and combinations of dots and dashes refer
    to different values of the mass ratio $q=M_\text{NS}/M_\text{BH}$,
    as indicated in the legend of the bottom right panel. The thick,
    black, continuous line marks the boundary between systems with
    $M_\text{b,torus}=0$ and ones with $M_\text{b,torus}>0$: these are
    located below and above the line, respectively. The
    $f_{220}^\text{QNM}$ curves are gray above this line to indicate
    that the excitation of the quasinormal mode is
    unlikely. \label{FIG:f220}}
\end{figure*}

We may now use our model to explore the space of parameters of BH-NS
systems systematically and to study the properties of the BH remnant
($a_\text{f}$, $\theta_\text{f}$, and its $l=m=2$, $n=0$ quasinormal
mode frequency). We do so by varying
\ben[~~~~(i)]
\item the binary mass ratio $Q$ between $2$ and $10$;
\item the initial dimensionless spin parameter of the BH,
  $a_\text{i}$, between $0$ to $0.99$;
\item the angle the BH spin initially forms with respect to the
  orbital angular momentum, $\theta_\text{i}$, from $0^\circ$ to
  $180^\circ$;
\item the NS mass, between $1.2M_\odot$ and $2.0M_\odot$, compatibly
  with the measurement reported in~\cite{Antoniadis2013};
\item the NS EOS.\footnote{Further details on the EOS are discussed in
    the Appendix of paper I.} More specifically, we quote most
  results for the APR2 EOS~\cite{Akmal1997,Akmal1998a}, since this is
  the most complete nuclear many-body study to date; additionally,
  when considering EOS-dependent effects (see Fig.\,\ref{FIG:f220}),
  we use the WFF1 EOS~\cite{Wiringa88} and the PS
  EOS~\cite{PandharipandeSmith:1975} as representatives of the softest
  and stiffest possible EOS, yielding the most and least compact NSs,
  respectively.
\een
We wish to remark that the model is untested in the $a_\text{i}\sim
0.99$ regime, where studies have shown that BH-NS systems behave very
nonlinearly~\cite{Lovelace2013} and where a basic assumption behind
the $M_\text{b,torus}$ fit of~\cite{Foucart2012} may break down.

\begin{figure*}[htb]
  \begin{tabular*}{\textwidth}{c@{\extracolsep{\fill}}c}
  \includegraphics[scale=1.3,clip=true]{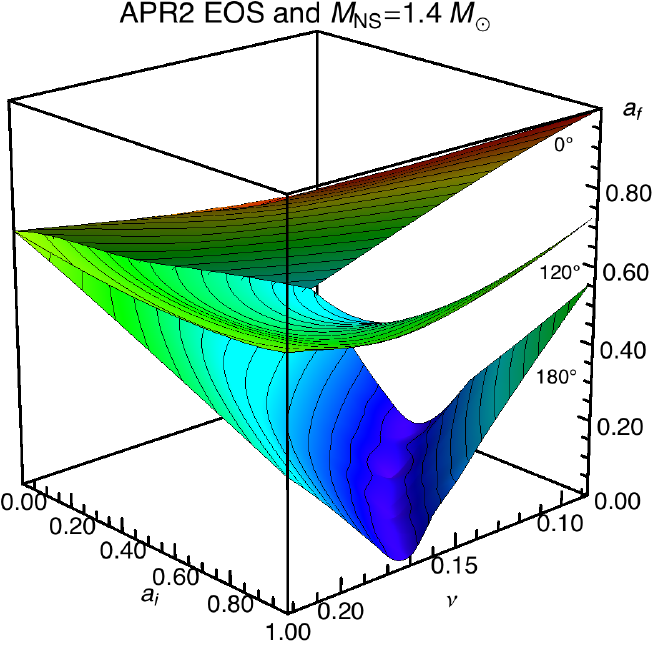}~~~
  &
  ~~~\includegraphics[scale=1.3,clip=true]{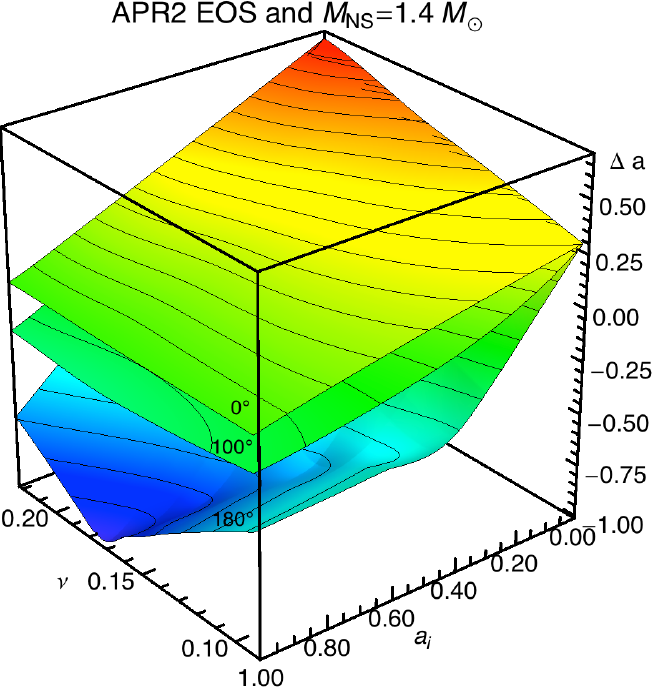}
  \\
  &
  \\
  \includegraphics[scale=1.3,clip=true]{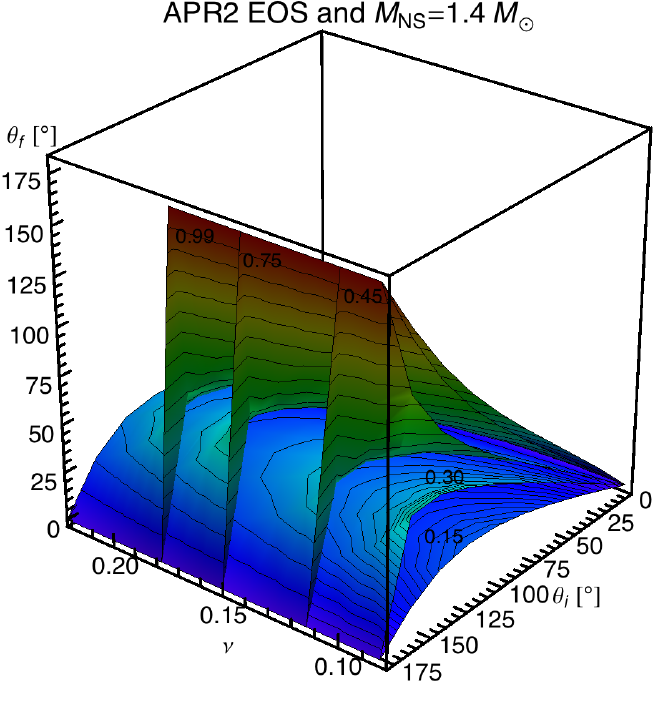}~~~
  &  
  ~~~\includegraphics[scale=1.3,clip=true]{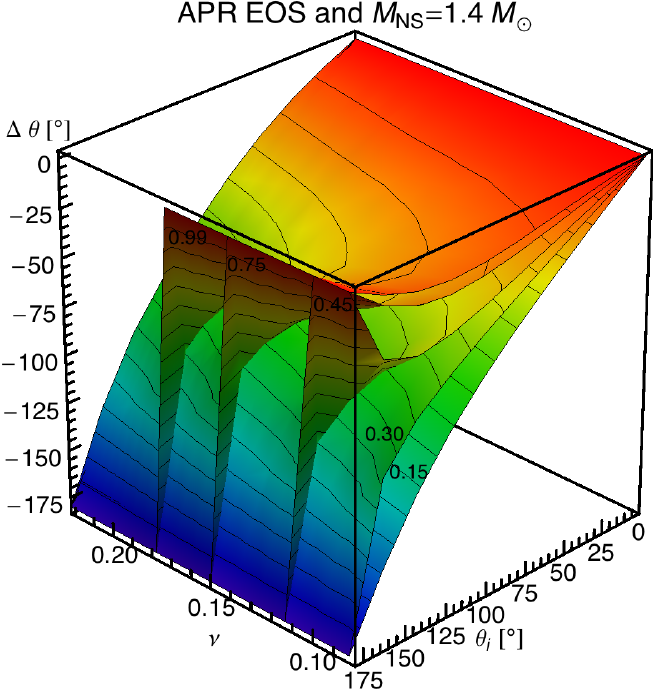}
  \end{tabular*}
  \caption{(Color online) APR2 EOS and $M_\text{NS}=1.4M_\odot$.
    Top left: $a_\text{f}$ as a function of $a_\text{i}$ and $\nu$
    for $\theta_\text{i}=\{0^\circ,120^\circ,180^\circ\}$ (as indicated by
    the labels on the surfaces).
    Top right: $\Delta a=a_\text{f}-a_\text{i}$ as a function of $a_\text{i}$
    and $\nu$ for $\theta_\text{i}=\{0^\circ,100^\circ,180^\circ\}$ (as indicated
    by the labels on the surfaces).
    Bottom left: $\theta_\text{f}$ as a function of $\theta_\text{i}$ and
    $\nu$ for $a_\text{f}=\{0.15,0.3,0.45,0.75,0.99\}$ (as indicated by
    the labels on the surfaces).
    Bottom right: $\Delta\theta=\theta_\text{f}-\theta_\text{i}$ as a function
    of $\theta_\text{i}$ and $\nu$ for $a_\text{i}=\{0.15,0.3,0.45,0.75,0.99\}$
    (as indicated by the labels on the surfaces).
    \label{FIG:3D}}
\end{figure*}

A first noteworthy result is that, when spanning the space of
parameters just outlined, our model does not produce overspinning BHs
($a_\text{f}>1$), which were instead encountered, and discarded,
in~\cite{Stone2013}.

We begin our analysis by discussing the results collected in the four
panels of Fig.\,\ref{FIG:3D}, where $M_\text{NS}=1.4M_\odot$ and the
APR2 EOS is used. In the top panels, we show $a_\text{f}$ and $\Delta
a\equiv a_\text{f}-a_\text{i}$ as a function of the symmetric mass
ratio of the binary, $\nu$, and $a_\text{i}$ for specific values of
the initial tilt angle $\theta_\text{i}$; in the bottom panels, we
show $\theta_\text{f}$ and
$\Delta\theta\equiv\theta_\text{f}-\theta_\text{i}$ as a function of
$\nu$ and $\theta_\text{i}$ for specific values of $a_\text{i}$. In
the top left panel we consider $a_\text{f}$ for
$\theta_\text{i}=0^\circ, 120^\circ$, and $180^\circ$. Notice that the
three surfaces all intersect at $a_\text{i}=0$, as expected. With the
exception of this line, for any combination of $a_\text{i}$ and $\nu$,
$a_\text{f}$ is highest if $\theta_\text{i}=0^\circ$, i.e.~if the BH
spin angular momentum and the orbital angular are aligned. As
$\theta_\text{i}$ is increased, the values of $a_\text{f}$ decrease
progressively. In particular, the conditions for having $a_\text{f}=0$
may develop, as is evident for the $\theta_\text{i}=180^\circ$
surface. Recalling that $a_\text{i}$ and $a_\text{f}$ are positive,
the shape of the $\theta_\text{i}=180^\circ$ surface tells us that,
under certain combinations of the initial physical parameters, a spin
flip may take place in BH-NS binaries. This means that the initial and
final BH spin angular momenta may be antiparallel. For high values of
$a_\text{i}$, this happens for high symmetric mass ratios, that is,
when the values of the NS and BH mass are close, so that the angular
momentum accreted onto the BH is more ``effective'' in increasing the
BH spin angular momentum. Notice that for $a_\text{i}\lesssim 0.4$ a
spin flip always occurs in the range of $\nu$ we consider.

In the top right panel of Fig.\,\ref{FIG:3D} we look at our results in
terms of $\Delta a$. This time, we show the surface for
$\theta_\text{i}=100^\circ$ instead of $120^\circ$, so that the
results for $\theta_\text{i}=180^\circ$ are visible more
clearly. Again, we see that the more the BH spin angular momentum is
initially aligned to the orbital angular momentum, the more this is
transferred to the BH, as expected, so that the highest values of
$\Delta a$ are obtained when $\theta_\text{i}=0^\circ$. The highest of
these, in particular, is encountered when the BH is initially
nonspinning and $\nu=2/9$, which corresponds to the lowest value of
$Q$ that we use. Notice, once more, the effects of spin flip on the
shape of the $\theta_\text{i}=180^\circ$ surface.

In the remaining two panels of Fig.\,\ref{FIG:3D} we show
$\theta_\text{f}$ and $\Delta\theta=\theta_\text{f}-\theta_\text{i}$
as a function of $\nu$ and $\theta_\text{i}$ for $a_\text{i}=0.15,
0.30, 0.45, 0.75,$ and $0.99$. For $a_\text{i}<0.30$,
$\theta_\text{f}\lesssim 50^\circ$, no matter what $\th_\text{i}$ is
considered. When $a_\text{i}$ is increased, however, higher and higher
angles $\theta_\text{f}$ are achieved for any given $\nu$ and
$\theta_\text{i}$. At $\theta_\text{i}=180^\circ$, this behavior
culminates in spin flip and, with the right combinations of
$a_\text{i}$ and $\nu$, $\theta_\text{f}$ can be $0^\circ$. As the
plots show, the critical value of the mass ratio for the onset of spin
flip increases with $a_\text{i}$. This means that as $a_\text{i}$
grows it is increasingly harder to flip the spin of the BH. On the
other hand, $\theta_\text{f}=0^\circ$ for any value of the symmetric
mass ratio when $a_\text{i}$ is low.

Knowing the mass and spin of the BH remnant, we may determine its
$l=2$, $m=2$, $n=0$ quasinormal mode (QNM) frequency
$f_{220}^\text{QNM}$. This is done via the fit provided
in~\cite{Berti:2009kk}. Our results for $f_{220}^\text{QNM}$ are
reported in the panels of Fig.\,\ref{FIG:f220}. The mass of the NS in
the BH-NS binary system is taken to be $M_\text{NS}=1.4M_\odot$ in all
panels. The initial dimensionless BH spin parameter is progressively
increased from top to bottom. The WFF1 and PS EOS are used in the left
and right panels, respectively. As in paper I, this choice is
motivated by the fact that these EOSs yield the most and least compact
NSs, respectively. In each panel, we report $f_{220}^\text{QNM}$ as a
function of $\theta_\text{i}$ for $q\equiv
Q^{-1}=M_\text{NS}/M_\text{BH}=0.1,0.2,0.3,0.4$, and
$0.5$.\footnote{This different definition of the mass ratio is used in
  order to obtain evenly spaced curves in the panels of
  Fig.\,\ref{FIG:f220} when varying the mass ratio by a fix amount.}
Physically, a decrease in the BH mass and an increase in the BH spin
both cause the QNM frequency value to increase. It is therefore not
surprising to observe that, overall, the QNM frequency increases as
$q$ and/or $a_\text{i}$ grow. A second characteristic all panels
share, is that, all else being fixed, increasing the initial tilt
angle $\theta_\text{i}$ causes $f_{220}^\text{QNM}$ to decrease. Once
again, this is due to the fact that the spin of the BH remnant
diminishes as $\theta_\text{i}$ grows. A third general behavior we
observe is that, given an NS EOS and an initial BH spin parameter, the
range of values spanned by $f_{220}^\text{QNM}$ as $q$ runs from $0.1$
to $0.5$ narrows as $\theta_\text{i}$ increases.

Within the space of parameters we consider, $f_{220}^\text{QNM}$
varies between $\sim 1$kHz and $6.5$kHz. A ringdown occurring at the
lower end of this range may be detected by second generation
gravitational-wave detectors. On the other hand, third generation
detectors with enhanced sensitivity in the high-frequency regime, like
the Einstein Telescope, will be necessary to access the possible
ringdown of systems yielding $f_{220}^\text{QNM}\gtrsim 2$Hz, unless
these are particularly close to Earth, obviously.

A key point is, naturally, whether the ringdown occurs or not, that
is, whether an oscillation at $f_{220}^\text{QNM}$ is excited or
not. A simple criterion to estimate whether this is the case or not
may be established by considering the mass $M_\text{b,torus}$ of the
material that is not accreted onto the BH. If $M_\text{b,torus}>0$,
the NS is tidally disrupted, (a part of) its matter accretes
incoherently onto the BH, and the ringdown of the BH is unlikely to be
excited. When $M_\text{b,torus}=0$ the accretion is more coherent ---
the majority of these mergers end with a plunge --- and therefore the
BH ringdown may be excited.\footnote{See~\cite{Pannarale2013a} for a
  more detailed discussion.} In the panels of Fig.\,\ref{FIG:f220}, a
black, thick, continuous line divides systems with
$M_\text{b,torus}>0$ from ones with $M_\text{b,torus}=0$: the former
are above the line, and the latter are below. The $f_{220}^\text{QNM}$
curves are colored when $M_\text{b,torus}=0$, that is, when the BH
ringdown is expected to be excited and, therefore, imprinted in the
emitted gravitational radiation. Notice that the EOS affects the
location of the boundary between systems such that
$M_\text{b,torus}=0$ and ones that yield $M_\text{b,torus}>0$, and
thus the portion of the parameter space in which a BH ringdown may
take place.

Results like the ones reported in Fig.\,\ref{FIG:f220} lead to a
number of considerations regarding GW detection and astrophysics. In
summary, our expectations in terms of the absence or presence of BH
remnant ringdown in the GW signal emitted by a coalescing compact
binary (i) may be used in guiding GW modeling and detection efforts
and (ii) are a valuable source of astrophysical information when
combined with detected GW signals. Let us illustrate these two
statements by means of some examples.
\bi
\item Point-particle GW templates are sufficient to detect BH-NS
  inspirals~\cite{Pannarale2011}. Figure \ref{FIG:f220} now tells us
  more: it shows where in the parameter space of BH-NS binaries a
  search may benefit from the use of \emph{BH-BH} GW templates beyond
  the inspiral regime, and where it would not. For example, and with
  the exception of highly spinning BH cases, BH-NS systems with mass
  ratios of $1:10$ are expected to behave like BH-BH binaries and to
  exhibit the ringdown of the BH remnant, so that the use of complete
  inspiral-merger-ringdown (IMR) binary BH waveforms would be
  favorable. On the other hand, searches for BH-NS systems with
  moderate tilt angle and low BH mass may be limited to the inspiral
  stage and employ point-particle templates; alternatively, they could
  profit from the construction of \emph{BH-NS} IMR templates.
\item Suppose the GW emission of a canonical BH-NS system with mass
  ratio $1:7$ is detected. If, say, the initial BH spin and the
  orbital angular momentum are roughly aligned
  ($\theta_\text{i}\lesssim 25^\circ$) and $a_\text{i}\sim 0.8$, the
  lack of BH remnant ringdown in the detected GW may rule out a BH
  binary with equivalent physical parameters as a possible source, and
  it would rule out soft EOSs. On the other hand, the repeated
  observation of systems of this kind exhibiting ringing BH remnants
  would allow us to place upper bounds on the NS EOS stiffness.
\item Suppose, instead, that the GW emission of a compact binary
  system with mass ratio between $1:4$ to $1:2$ is detected. If, say,
  the initial spin angular momentum of the secondary is negligible,
  and the initial spin angular momentum of the primary is roughly
  aligned with the orbital angular momentum ($\lesssim 25^\circ$) and
  is such that the dimensionless spin parameter is $\sim 0.8$, then
  the absence of a BH remnant ringdown would tell us that this is a
  BH-NS system with a particularly low total mass, and not a BH-BH
  system. This, in turn, would have implications in terms of
  population synthesis, as would the repeated observations of binaries
  with these properties and manifest BH remnant ringdown.
\ei

In order to be feasible, the scenarios discussed necessitate further
investigation in terms of data analysis, of course. In particular,
determining whether the ringdown of the BH remnant is present in a
detected GW requires a rigorous formulation. Our discussion, however,
illustrates that our model may be used to break degeneracies between
BH-BH and BH-NS systems. This in turn could improve our understanding
of compact binary formation and help us constrain the NS EOS.

\section{Conclusions and Remarks}\label{sec:conclusions}
In this paper we presented a model that allows us for the first time
to accurately determine the final spin parameter, $a_\text{f}$, and
mass, $M_\text{f}$, of the BH remnant of BH-NS coalescing binaries,
with arbitrary mass ratio, NS mass, NS cold, barotropic EOS, and BH
spin magnitude and orientation. This work generalizes the model
presented in paper I, which was limited to mixed binaries with aligned
BH spin and angular momentum. It is based on the Buonanno, Kidder, and
Lehner model for the final spin of binary BH
mergers~\cite{Buonanno:07b}. Our approach relies on a fit to the mass
of the torus remnant of aligned BH-NS mergers~\cite{Foucart2012} and
on its use for BH-NS mergers with nonparallel BH spin and orbital
angular momentum, a possibility just recently pointed
out~\cite{Stone2013,Foucart2013a}.

Exploring the BH-NS parameter space via numerical-relativity
simulations is desirable but in practice very time and resource
consuming. Further, results for misaligned initial BH spin
configurations were so far presented by the Caltech-Cornell-CITA
Collaboration only\cite{Foucart2010,Foucart2013a}. While this means
there is just a handful of possible tests that allow us to validate
the model discussed in this paper for binaries with inclined initial
BH spin angular momentum, it also means that we are now able to study
certain aspects of BH-NS coalescing systems for the first time.

The model was tested by comparing its predictions to the recent seven
numerical-relativity simulations of BH-NS mergers with inclined
initial BH spin angular momentum. The agreement between our
predictions and the numerical results is at the level of the tests
performed in paper I for systems with aligned BH spin: the maximum
absolute error on the final BH spin parameter is $0.02$, while the
maximum relative error on the final BH gravitational mass is $1$\%
when using the numerical-relativity result for the mass of the torus
remnant and $2$\% when using the prediction of~\cite{Foucart2012} for
the same quantity. Additionally, in
Eq.\,(\ref{eq:disk-tilt-constraints}) we were able to establish how
the final inclination angle $\theta_\text{f}$, which is another
natural output of the model [see
Eqs.\,(\ref{eq:LorbAccr})-(\ref{eq:model-tilted-2})], may be used to
constrain the misalignment angle between the disk possibly formed in
the merger and the final BH spin angular momentum vector. The
importance of this angle lies in the fact that it is directly linked
to observational signatures that may help us distinguish between SGRBs
of BH-NS origin and ones of NS-NS origin~\cite{Stone2013}. Finally, we
compared our model and a PN-based model that predicts the
dimensionless spin parameter of the BH remnant~\cite{Stone2013}. The
former was seen to be more accurate, especially when low BH masses are
involved, particularly in combination with high initial spin
magnitudes. Further, and as opposed to the PN-based model, our model
is well behaved, in the sense that it does not produce overspinning BH
remnants.

The approach presented and tested in the first part of this work was
used to span the space of parameters consisting of the binary mass
ratio, the NS mass, the NS EOS, and the BH initial spin magnitude and
inclination. The model and results presented here improve our
understanding of BH-NS systems throughout their space of parameters
and may in principle be used in a number of ways. \emph{Prior to
  detection}, they may be exploited to determine the best strategy to
adopt when targeting GWs emitted by BH-NS systems in a specific
portion of the space of parameters (utilizing point-particle inspirals
vs~BH-BH IMR templates vs~BH-NS IMR templates). Additionally, they
show us where in the space of parameters the development of BH-NS IMR
templates is especially important (roughly speaking, low total masses,
moderate to high BH spins, and low to moderate, or even high if
considering stiff NS EOSs, initial BH spin tilt angle): this is where
improving the physics and the number of available numerical-relativity
simulations would be particularly beneficial. At the same time, this
also indicates the direction in which recent work on the interface
between numerical and analytical modeling of GWs should be
extended~\cite{Pannarale2013a, Lackey2013}, possibly using the model
presented in this paper. The results also, inform us of where BH-NS GW
detection directly benefits from the development of full BH-BH
templates (roughly speaking, this is the rest of the parameter space,
in line with the conclusions of~\cite{Foucart2013b}). Naturally, the
comparison between BH-NS and BH-BH merger simulations and GW templates
is important throughout the space of parameters, but it is
particularly so on the border between the two areas of the parameter
space that we identified. Strong efforts in this direction recently
appeared in~\cite{Foucart2013b}. \emph{Following detection}, knowledge
of the properties of the BH remnants of BH-NS systems may allow us to
break degeneracies and aid us in deciding whether a GW from a
coalescing binary was emitted by a BH-NS system or not, in placing
constraints on the NS EOS, and in constraining compact binary
formation scenarios. Improving our understanding and predictions of
the GW cutoff frequency, as recently done for nonspinning BH-NS
systems~\cite{Pannarale2013a}, is especially important in the context
of pinning down the NS EOS.

By allowing for generic initial BH spin configurations, the
formulation of the model presented in this paper completes one of the
possible extensions outlined in paper I. The original version of the
model has proven to be useful in developing phenomenological waveforms
for nonspinning BH-NS systems~\cite{Pannarale2013a}. In the future, we
plan to first extend the waveforms of~\cite{Pannarale2013a} to cases
with a spinning BH with aligned spin, and then to consider initially
misaligned BH spin
configurations~\cite{Lundgren2013b,Harry2013,Hannam2013b}. As for
paper I, we reiterate our intention to test and upgrade the model as
more numerical-relativity simulation results become available.

\section*{Acknowledgments}
It is a pleasure to thank Andrea Maselli and Frank Ohme for carefully
reading the manuscript and providing useful comments. The author
wishes to thank the anonymous referee for useful remarks which helped
improve the paper. This work was supported by STFC Grant
No.~ST/L000342/1 and by DFG grant SFB/Transregio~7.

\bibliographystyle{apsrev4-1-noeprint}
\bibliography{aeireferences}
\end{document}